# Voice-controlled Debugging of Spreadsheets

Derek Flood, Kevin Mc Daid,
Dundalk Institute of Technology, Dublin Road, Dundalk
derek.flood@dkit.ie, kevin.mcdaid@dkit.ie

**ABSTRACT**

*Developments in Mobile Computing are putting pressure on the software industry to research new modes of interaction that do not rely on the traditional keyboard and mouse combination. Computer users suffering from Repetitive Strain Injury also seek an alternative to keyboard and mouse devices to reduce suffering in wrist and finger joints. Voice-control is an alternative approach to spreadsheet development and debugging that has been researched and used successfully in other domains. While voice-control technology for spreadsheets is available its effectiveness has not been investigated. This study is the first to compare the performance of a set of expert spreadsheet developers that debugged a spreadsheet using voice-control technology and another set that debugged the same spreadsheet using keyboard and mouse. The study showed that voice, despite its advantages, proved to be slower and less accurate. However, it also revealed ways in which the technology might be improved to redress this imbalance.*

## 1. INTRODUCTION

It is not convenient or efficient for all spreadsheet users at all times to control spreadsheet technology through the usual keyboard and mouse combination. With the move towards mobile and smaller battery powered computer devices and the growing number of sufferers of Repetitive Strain Injury (RSI), alternatives to the keyboard and mouse interface systems are receiving much attention. As most mobile computers feature a built in microphone, speech interface to applications is an obvious technological option. Speech interfaces have the additional benefit in that they allow users communicate in a natural way. This paper explores the use of a speech interface for spreadsheet debugging.

Spreadsheets are becoming more complex in nature and as such the risk of errors remaining in spreadsheets is increasing. Often spreadsheets are used as the basis for important business decisions and therefore it is important that they are error free. Unfortunately this is not the case, with many studies recording an unacceptably high level of errors in spreadsheets[Brown 1987, Panko 1998]. One such example caused stock values at Shurgard Inc to fall sharply after two employees were overpaid by $700,000 [Fisher, 2006].

In order to ensure that errors like this do not occur, spreadsheets should be thoroughly checked and debugged. This process involves reviewing the spreadsheet to ensure that no errors exist and if any errors do exist then they are repaired. The removal of these errors is known in software development as debugging. Section 2 of this paper highlights a number of issues with spreadsheet development.

There are existing speech-based technologies for the development and debugging of spreadsheets, the most popular being Dragon NaturallySpeaking developed by Nuance [nuance, 2007]. These systems allow users to control the spreadsheet application through voice commands, providing users with shortcut phrases for navigation, data and formula entry, and formatting. Section 3 reviews the state of the art in voice-control technology.

This work evaluates the effectiveness of current state of the art technology for voice control of spreadsheet debugging. We have conducted a trial in which a set of users debug a spreadsheet through voice and compare the results with users who preformed the same task using traditional keyboard/mouse control. We found that participants that used voice technology found on average 15 % fewer errors and took almost twice as long to complete the task. There were also significant differences in the time to navigate through the spreadsheet and the time to implement changes. Section 4 details the trial and the environment. An analysis of both the quantitative data, gathered through an experiment, and the qualitative data gathered through a structured questionnaire-based interview is presented in Section 5.

Section 6 concludes the paper with a summary of proposed future work in the area.

## 2. SPREADSHEETS

Spreadsheets are an important decision making tool. They help companies make multi-million euro decisions on a daily basis. The financial district within the city of London is responsible for 3% of the United Kingdoms Gross Domestic Product (GPD). Within this one square mile, spreadsheets have been described as the primary front line tool of analysis [Croll, G., 2005]. Despite the importance of the decisions being made based on these spreadsheets, there is very little done in practice to ensure their quality.

A number of researchers and consultancy firms have looked at the quality of spreadsheets. One consulting firm in England, Coopers and Lybrand, found that 90% of spreadsheets with over 150 rows contained errors [Panko, 1998]. These errors range in severity from simple spelling errors to complicated formula errors.

A range of techniques have been proposed to improve the development of spreadsheets. A number of the initiatives focus on the importance of development methodologies similar to existing rigid document driven lifecycle processes in place for software development. More Flexible approaches have been adapted from agile software development methodologies. [O' Beirne, 2002; Rust, A., Bishop, B., Mc Daid, K., 2006].

Interesting research is ongoing in the use of Assertions [Burnett, M., Rothermel, G., Cook, C., 2006]. This method allows developers to place restrictions on the value of a given cell. The cell may be just a number or may include a formula. When a cell contains a formula the application calculates a value that the cell should be based on the assertions placed on the input cells. If there is a conflict between the value within the cell and the assertion placed on the cell then the user is notified as to a potential error.

## 3. VOICE COMMAND TECHNOLOGY

Voice command technology refers to the use of speech and voice recognition to enter information into a computer system. The system listens for certain key words or commands and performs a predefined action upon hearing these commands. This technology is not new. However, with the increase in computer power in recent years the technology has improved to the point that accuracy of the tools is advertised to be of the order of 99% [Nuance, 2006].

Improvements in technology have also allowed computers to shrink in size so that a computing device can fit easily into your pocket. For these portable computers, traditional input methods are no longer practical and alternatives must be explored. Voice command technology is the obvious candidate for control of such devices.

Voice command technology could also help address the issue of Repetitive Strain Injury (RSI) in the work place. The pain of this injury can prohibit sufferers from using a keyboard. In the UK 1 in 50 of all workers have reported an RSI condition [RSIA, 2007]. As a result every day six people in the UK leave their jobs[RSIA, 2007]. While voice technology is not the complete solution, it could remove the need for sufferers to use a keyboard while performing most computer related tasks.

Voice command technology has been used for creating and modifying text documents [Begel, 2005]. and has had widespread uptake. While Tony Stockman [2005] has looked at the sonification of spreadsheets, to the best of the authors' knowledge there has been no research published in the area of voice-controlled spreadsheet development.

With the increased interest in the area of voice technology, there has been a flurry of new applications across multiple domains. Andrew Begel [2005] has looked at using voice technology to allow developers to develop java programmes through speech with the development of *Spoken Java*. Spoken Java is a naturally verbalizable alternative to java. It allows developers to speak code in a natural way, and have the java source code generated based on their statements.

Voice control of in-car navigation systems has also been investigated. While driving in a foreign country drivers frequently have to consult maps in order to find where they are going. Coletti et al [2003] have developed an in-car assistance where drivers can ask questions in a natural language and the system will provide the driver with the results through voice. This way the driver does not have to take their eyes off the road. The system allows the driver not only to get directions but can also allow them to make hotel and restaurant reservations.

Voice commands have also been used by Wang [2006] to develop a 3-D animation application. As with traditional movie sets, the director tells the actor what they need them to do and the actors perform as instructed. Wang has developed an application that allows a 3-D animator to act like a director and tell the objects on screen what to do and the objects will perform as instructed. The system is based on voice commands where the director issues the commands and the actors and objects behave based on predefined actions for the commands.

There are many voice recognition tools available; the best of these tools is Dragon NaturallySpeaking. Dragon NaturallySpeaking began in the early 1980's as DragonDictate [wikipedia, 2007]. Since then it has passed through 4 companies and is currently being sold by ScanSoft Inc. There are currently 5 versions of this popular dictation software on the market. The most basic of these, Standard edition, features support for Microsoft© Word and web browsing through Microsoft© Internet Explorer. The next edition, the preferred edition, includes support for Microsoft© Excel. The professional Edition builds on this again, by offering support for all Microsoft© Office applications. There are also two special editions, Legal edition and Medical edition, which provide special recognition for specialised legal and medical terms [nuance, 2007].

Microsoft Speech API is a competing voice recognition technology. It has two modes of operation, "command and control" and "dictation". The "command and control" mode allows users to use predefined commands. While in this mode the software will only recognise words that have been defined as commands. While in "dictation" mode the software will interpret words and phrases as text and will ignore any commands that may be associated with these words. Users are required to manually switch between the two modes, which is often a source of frustration for new users. Dragon on the other hand integrates both modes.

The voice recognition software evaluated to date tends to be slow to process input. In order to recognise input the voice recognition software takes time to interpret the sound waves and understand what the user said. While the time lag is only of the order of seconds, any application would involve the processing of a large number of commands which can on occasion result in a significant net delay.

## 4. TRIAL AND RESULTS

The aim of the trial is to investigate the efficiency of voice control technology for debugging spreadsheets. The evaluation will focus on looking at the following three tasks, numeric input to cells, formula input and modification, and navigation through the workbook.

Three participants with significant experience in the development of spreadsheets partook in the trial. None of the participants had any prior experience using Dragon NaturallySpeaking or any voice recognition software. For the duration of the trial they where placed in a quiet room to minimise the amount of background noise and thus maximise the effectiveness of the voice recognition system. In the room was one additional observer, the first author, to answer any questions participants may have had during the trial.

The trial was run on a Dell Notebook D610 with an Intel Pentium M 1.60 GHz , 1.0 GB of RAM running Windows XP Service Pack 2. For the voice recognition software Dragon NaturallySpeaking 9 Preferred Edition was used. A Creative HS – 600 headset provided the audio input.

The session included three elements and took approximately 2 hours. The first element asked participants to read a set of predefined text so that Dragon NaturallySpeaking could be calibrated for their voice. This took approximately ten minutes with no significant difficulties experienced by any of the participants.

The second element of the trial allowed participants to become familiar with the commands available to them within the spreadsheet environment. Participants were given a basic task involving input of text, data and basic formulas. This task was not assessed in any way and was designed purely to give the participants the opportunity to familiarise themselves with the technology and the range of commands. Participants were encouraged to experiment with the various inbuilt commands. Forty minutes was assigned to this element of the trial.

The final element of the trial included the experiment central to the study. Participants were given a pre-designed spreadsheet containing forty-two errors, which can be divided into 4 categories, Clerical, Formula Input, Rule Violation and Data Input, and were asked to find and repair as many errors as possible. Participants were only allowed to use the voice recognition technology and were not allowed to use the keyboard or mouse. Participants were handed a list of rules and an explanation of what was required of the various parts of the spreadsheet. They were not however told the types of errors to look for. There was no time limit set for this activity with participants finishing when they felt they had completed the task. Note that the workbook contained three sheets named, "Payroll", "Office Expenses" and "Projections".

This experiment is identical in structure to one developed by Bishop and McDaid [2007] in which 13 professionals completed the task. The analysis section will compare the performance of the three voice interface subjects with the professional subjects that used keyboard and mouse technology in the Bishop and McDaid Study.

It is important to note that technology was used to record the time of all selection and editing actions in the spreadsheet. As this was also the case for the study of keyboard users a more detailed comparison of the results could be made. Furthermore, audio recordings of the participants' commands throughout the experiment were kept using Easy MP3 sound recorder.

Finally, upon completion of the experiment participants were interviewed through a structured questionnaire. This was used to establish their views on the technology and the study.

## 5. ANALYSIS OF RESULTS

|  | Average Performance | Average Time |
|---|---|---|
| **Voice Group** | 57 % | 64 Minutes |
| **Keyboard Group** | 72 % | 28 Minutes |

Table 1 Overall Performance

### 5.1 Overall

It was found that, compared with 13 professionals who used keyboard/mouse technology, participants of this study took significantly longer. The average time for the voice participants was almost 64 minutes compared to 28 minutes for traditional users. The fastest voice participant was just over 55 minutes, almost twice as long as traditional input. These time differences could have serious consequences within a business environment.

Despite the longer time, it was found that voice participants under performed when compared with their keyboard counterparts, with an average of 57% of the errors corrected whereas keyboard users found and corrected 72% of the errors. The results for the three subjects where 45%, 50% and 76%.

Although the participants in the voice study differed from those in the keyboard/mouse study all subjects had extensive experience of spreadsheet development and would have been expected to perform to a similar standard in the debugging experiment.

|  | Numeric Input | Formula Input |
|---|---|---|
| **Voice Group** | 18.1 | 73.5 |
| **Keyboard Group** | 5.5 | 7.3 |

Table 2: Average Input Times in Seconds

### 5.2 Editing Formulas

The analysis examined in detail the time to navigate and edit the cells of the workbook. It was found that most of the time was spent editing formulas. With a keyboard this process is fairly straightforward, however when using voice technology this proved to be quite difficult. It was found that on average voice users took 73 seconds to edit a formula whereas the keyboard group were able to perform the same tasks in an average of 7 seconds. This was based on a set of 9 errors that where edited by all voice subjects and the average value for the cell was compared with the average for those professionals that edited the cell. The average time to edit the nine cells is given in Table 3. This striking difference comes from the method by which voice users had to edit the formulas. In order to move a single character to the left within the formula, the voice users need to say "Move Left" and wait for the recognition engine to recognise what they have said and then to perform the move. In contrast keyboard users just need to hit the left arrow key.

In cell I14 of the payroll worksheet the voice users took on average 290 seconds. The main reason for this large time is that one participant took almost 7 minutes to edit this cell. The issue was with the word "sum" and the letter "m" which was repeatedly interpreted by the recognition software as "n".

| Cell | Voice Group | Keyboard Group |
|---|---|---|
| **Payroll F10** | 30.4 | 4.6 |
| **Payroll G16** | 19.2 | 5.5 |
| **Payroll I10** | 32.2 | 6.1 |
| **Payroll I14** | 290.3 | 6.2 |
| **Office Expenses F8** | 6.0 | 4.4 |
| **Office Expenses F5** | 23.7 | 9.7 |
| **Office Expenses F20** | 154.7 | 13.7 |
| **Projections B17** | 87.4 | 7.6 |
| **Projections G22** | 18.1 | 7.6 |

**Table 3: Average time to edit formulas in seconds**

### 5.3 Entering Data

Participants using voice-control technology found it easier to enter numeric data as opposed to formulas. The time it took to enter numbers on average for voice control subjects was found to be 18 seconds in comparison with keyboard users who took on average almost 6 seconds. This was based on 6 cells and the complete set of values is shown in Table 4.

|  | Voice Group | Keyboard Group |
|---|---|---|
| **Payroll C13** | 47.0 | 8.6 |
| **Payroll D10** | 20.1 | 7.9 |
| **Payroll D11** | 7.9 | 7.6 |
| **Payroll E14** | 23.1 | 1.6 |
| **Office Expenses D16** | 6.4 | 5.1 |
| **Office Expenses D17** | 4.4 | 2.4 |

**Table 4: Average time to enter numbers in seconds**

### 5.4 Navigation

Another important aspect to spreadsheet debugging is navigation of the spreadsheet. It was found that using voice technology proved to be slower. In particular we examined the time taken to traverse five cells containing a copied formula without errors. An initial investigation showed that voice technology subjects take significantly longer to traverse spreadsheets. We are also exploring the time to navigate between spreadsheets. Preliminary studies indicate that it is extremely time consuming to switch between sheets when attempting to investigate formulas linked to cells in other sheets.

### 5.5 Workarounds

When assessing the efficiency of voice controlled technology it should be stated that in certain situations users were unable to complete a task in the usual way and used workarounds to complete the task. An example of this relates to one of the errors within the spreadsheet where the word "Mar" was misspelled as "Mac". One particular user was having trouble getting the recognition software to recognise the letter "r", as it repeatedly produced the number "4". In the end the user abandoned trying to spell the word and instead chose to copy the word from elsewhere and pasted it into the cell.

These results show that, at present, there are real issues with the use of voice technology to debug spreadsheets.

After they had completed the task, users were asked a series of questions to determine their experience of using the voice recognition software in the trial. Two out of the three users felt that using voice commands distracted them from the task at hand. However one user found that using voice made him concentrate more so that they would not have to redo part of the spreadsheet again. All of the participants agreed that learning the voice commands was relatively easy and intuitive.

### 6. CONCLUSIONS

Voice recognition technology can be used to debug spreadsheets, however it was found that it is associated with poorer performance and takes more time than traditional keyboard and mouse entry. By providing users with a spreadsheet containing errors and asking them to find and repair them, one group using voice and one group using traditional input methods, it was found that voice recognition users took nearly twice as long and found 15% less errors.

From what we have learned, we do not believe that voice technology in its current form is a very suitable means for spreadsheet debugging. However, we feel that with certain improvements it may in the future be possible to use the technology.

In particular we believe that the system could be improved through the addition of an intelligent navigation system that suggests the most useful navigational actions for the user. A recommender system [McCarey, 2004] may be used to identify the most useful actions. These actions will not be automatically executed; instead they will be displayed so that the user can if they wish take advantage of them through a simple set of voice commands. Another possible enhancement could be the development of a more intuitive formula input system. This system could use Begel's Spoken Java [Begel, 2005] as a starting point.

Another area of interest is kinaesthetic memory which relates to the memory in relation to space. In [Tan, 2002] the authors compare the use of touch screen technology with mouse technology. They found that users using the touch screen could remember more accurately than those using a mouse. It would be interesting to see how this might impact on voice controlled development of spreadsheets.

## 7. REFERENCES

Begel, A.,(2005) "Spoken Language Support for Software Development"

Bishop, B., McDaid, K.,(2007) "An Empirical Study of End-User Behaviour in Spreadsheet Debugging", Proceedings of 3$^{rd}$ Annual Work-in-Progress Meeting of the Psychology of Programming Interest Group

Brown, P. S., & Gould, J. D. (1987). An Experimental Study of People Creating Spreadsheets. ACM Transactions on Office Information Systems, 5(3), 258-272.

Burnett, M., Rothermel, G., Cook, C., (2006), End User Development, Kluwer Academic Publishers.

Coletti, P., Cristoforetti, L., Matassoni, M., Omologo, M., Svaizer, P., Geutner, P., Steffens, F., (2003)"A Speech Driven in-car assistance system", Proceedings of IEEE IV2003 Intelligent Vehicles Symposium, Columbus (Ohio), USA, June 9-11.

Croll, G., (2005) "The Importance of spreadsheets in the city of London", The EuSpRIG – European Spreadsheet Risks Interest Group

Fisher, M., Rothermel, G., Brown, D., Cao, M., Cook, C., and Burnett, M.,(2006) "Integrating Automated Test Generation into the WYSIWYT Spreadsheet testing methodology", ACM Transactions on software Engineering and Methodologies,, Vol.15, No. 2, April, pages 150-194.

McCarey F, ó Cinnéide, M, and Kushmerick, N., "A Case Study on Recommending Reusable Software Components using Collaborative Filtering", Proceedings of the MSR workshop, 2004.

http://www.nuance.com/naturallyspeaking/preferred/ accessed 26$^{th}$ February 2007

http://www.nuance.com/naturallyspeaking, Accessed 27$^{th}$ February 2007

O' Beirne, P. (2002) "Agile Spreadsheet Development (ASD)" [online] Available: http://www.sysmod.com/agile.htm [5th March 2007].

Panko, R., Halverson, R., (1996) "Spreadsheets on Trial: A survey of Research on Spreadsheet Risk", Proceedings of the 29$^{th}$ Annual Hawaii International Conference on System Sciences.

Panko, R.,(1998) "What we know About Spreadsheet Errors", Journal of End User Computing 10, 2 (Spring), p15-21

Rust A., Bishop B., Mc Daid, K.,(2006) "Investigating the potential of Test-Driven Development for Spreadsheet Engineering", The proceedings of the European Spreadsheet Risks Interest Group

"RSI Facts and Figures", The Repetitive Strain Injury Association, http://www.rsi.org.uk/pdf/Facts_&_Figures.pdf accessed 26 February 2007.

Stockman, T., Hind, G., Frauenberger, C., "Interactive sonification of Spreadsheets", Proceedings of ICAD 05-Eleventh Meeting of the International conference on Auditory Display, 2005

Tan, D. S., Stefanucci, J. K., Proffitt, D.R., Pausch, R. "Kinesthetic Cues Aid Spatial Memory", (2002) *Extended Abstracts at CHI 2002 Conference on Human Factors in Computing Systems, pp. 806-807*

http://en.wikipedia.org/wiki/Dragon_NaturallySpeaking#History, Accessed 27th February 2007

Wang, Z., Van de Panne, M.,(2006) " "Walk to Here": A voice Driven Animation System", Eurographics / ACM SIGGRAPH Symposium on Computer Animation

Blank Page